\documentstyle[amssymb,aps,prl,twocolumn,epsf]{revtex}

\begin{document}
\draft

\flushbottom
\twocolumn[\hsize\textwidth\columnwidth\hsize\csname
@twocolumnfalse\endcsname
\title{Quantum Magnetic Impurities in Magnetically Ordered
Systems}
\author{A.~H.~Castro Neto$^1$, E.~Novais$^1$, L.~Borda$^{2,3}$, Gergely Zar\'and$^{2,4}$, and I.~Affleck$^1$}

\address
{$^1$ Department of Physics, Boston University, Boston, MA 02215\\
$^2$ Research Group of the Hungarian Academy of Sciences and
Theoretical Physics Department, TU Budapest, H-1521, Hungary\\
$^3$ Sektion Physik and Center for Nanoscience, LMU M\"unchen, 80333
M\"unchen, Theresienstr. 37, Germany\\
$^4$ Lyman Physics Laboratory, Harvard University, Cambridge, MA \\}

\date{\today }
\maketitle
\widetext\leftskip=1.9cm\rightskip=1.9cm\nointerlineskip\small

\begin{abstract}
\hspace*{2mm}
We discuss the problem of a spin $1/2$ impurity immersed in a spin $S$
magnetically ordered background. We show that the
problem maps onto a generalization of the dissipative two level system (DTLS)
with two independent heat baths, associated with
the Goldstone modes of the magnet, that couple to different 
components of the impurity spin operator. Using analytical 
perturbative renormalization group (RG) methods and accurate 
numerical renormalization group (NRG)
we show that contrary to other dissipative models there is
quantum frustration of decoherence and quasi-scaling even
in the strong coupling regime. We make predictions
for the behavior of the impurity magnetic susceptibility that can be measured
in nuclear magnetic resonance (NMR) experiments. 
Our results may also have relevance to quantum computation. 
\end{abstract}


] \narrowtext

Quantum impurity problems are characterized by a single quantum
mechanical degree of freedom coupled to a reservoir. These
are the simplest problems in physics that exhibit non-trivial 
many-body effects. Among them, the DTLS
\cite{leggett} plays a central role because it is related to a 
variety of different physical processes such as the 
dissipative tunneling of
a particle in a double well or the coupling of an Ising 
spin to a gapless fermionic environment, i.e, the anisotropic 
Kondo problem. The Kondo problem is one of the best understood
impurity problems and has been studied by a large variety of
methods. Its thermodynamic properties can be studied exactly 
via Bethe ansatz \cite{bethe-ansatz} and RG \cite{ayh}, 
its dynamic properties at low energies can be calculated 
via conformal field theory \cite{cft},
and many of its correlation functions can be obtained exactly 
\cite{saleur}. The problem of quantum impurities immersed in
magnetic media close to a quantum phase transition has also
attracted a lot of attention recently due to its possible
relevance to cuprates, heavy-fermions and organic materials
\cite{sengupta}. While most of the recent works focus on the
paramagnetic phase we concentrate on magnetically ordered
phases \cite{ferro,next}. 

In this paper we study a quantum impurity problem
of a different nature, namely, the problem of a single spin $1/2$
coupled to a $d$-dimensional magnetically ordered system with spin $S$. 
A possible application of our results can be found, for instance, 
in KMn$_{1-x}$Cu$_x$F$_3$ with $x \ll 1$ (Cu has spin $1/2$ and Mn has
$S=5/2$). KMnF$_3$ is a three dimensional (d=3) cubic quantum
Heisenberg antiferromagnet (QHAF) with exchange coupling $J$
between the Mn spins. This material 
orders in a N\'eel state and has well defined gapless magnon modes 
\cite{buyers}. The Mn spins interact with the Cu spin via an
exchange $J'$. In the past this problem was studied 
by a series of different theoretical techniques that usually
assume the impurity spin to be aligned with the
surrounding N\'eel state and/or $J' \approx J$ \cite{buyers,sachdev}. 
As we are going to show,
while this kind of approach is warranted at low
temperatures, it fails to describe the full quantum behavior
of the impurity at energy scales intermediate between $J'$ and $J$ ($J' \ll J$)
where the impurity
spin fluctuates strongly and gets mixed with the quantum fluctuations 
of the magnetic environment.
The problem at hand is similar to the Kondo effect of
magnetic impurities in metals where there is a 
many-body crossover from weak (paramagnetic) to strong coupling
(screened)
as the temperature is lowered below the Kondo temperature.
In our case, the screened state occurs when the impurity spin
fully aligns with the surrounding background below an energy
$T_A$. We are particularly interested in the anomalous
relaxation of the Cu ion spins that can be measured
by NMR. In particular, we calculate the frequency and
temperature dependence of the imaginary part of the 
transverse impurity susceptibility for the QHAF in $d=3$,  $\chi_{\perp}(\omega,T)$.

At long wavelengths and low energies the magnetic problem 
can be described by spin coherent state path integral in terms
of the Euclidean action ${\cal S}_E$ (we use units such that $\hbar=k_B=1$) 
with ${\cal S}_E={\cal S}_{WZ}+{\cal S}_M$ 
where ${\cal S}_{WZ}$ is the Wess-Zumino term that describes
the quantum dynamics for the impurity spin ${\bf S}$ and
\begin{eqnarray}
{\cal S}_M &=& \int d^{d+1} x_{\mu} 
\left\{\frac{1}{2 g} \left[
\left(\partial_{0} {\bf n}(x_{\mu})\right)^2 +
c^2 \left(\partial_{i} {\bf n}(x_{\mu})\right)^2 
\right] \right.
\nonumber
\\
&+& \left.
\delta^d(x_i)  \, \, \, {\bf n}(x_{\mu}) \cdot \lambda \cdot {\bf
  S}(x_0)\right\}  \, ,
\label{action}
\end{eqnarray}
where $x_{\mu}=(x_0=\tau,x_i)$ with $\mu=0,1,..,d$ 
is the space-time coordinate, $c = 2 \sqrt{d} J a S$ is
the spin-wave velocity, $g = c^2/\rho_s$ 
is the coupling constant, $\rho_s = J S^2 a^{2-d}$  
is the spin stiffness for the non-linear sigma model \cite{chn} 
described by the
vector field ${\bf n}$ ($a$ is the lattice spacing), 
$\lambda \propto J'$ is the matrix coupling between
the impurity spin and the spin environment. 
The action (\ref{action}) has to be supplemented by the local 
constraint ${\bf n}^2(x_{\mu}) =1$. 
In the ordered phase, we can write
${\bf n} \approx (\varphi_1,\varphi_2,1)$ 
with $|\varphi_a| \ll 1$. 
We will assume $\lambda=(\lambda_{1},\lambda_2,\Delta)$ and that 
$\Delta,\lambda_{1,2} \ll D_0 \approx J$ where $D_0$ is the
bare cut-off of the problem.
The action reduces to:
\begin{eqnarray}
{\cal S}_M &\approx& \int d^{d+1} x_{\mu} 
\left\{\frac{1}{2 g} 
\left[
\left(\partial_{0} \vec{\varphi}(x_{\mu})\right)^2 +
c^2 \left(\partial_{i} \vec{\varphi}(x_{\mu})\right)^2 
\right]
\right.
\nonumber
\\
&+& \left.
\delta^d(x_i)  \left[\Delta S_3 + \sum_{a=1,2} 
\lambda_a \varphi_a(x_{\mu}) S_a\right]\right\} \, , 
\label{action_order}
\end{eqnarray}
where $\vec{\varphi} = (\varphi_1,\varphi_2)$ represents the
two Goldstone modes of the antiferromagnet. 
Eq.(\ref{action_order}) describes a problem of two
free bosonic modes coupled to an impurity via its different
spin components. Notice that the
$\Delta$ coupling describes the molecular
Weiss field applied by the antiferromagnet. 
The other two terms have a very different meaning, they represent
the quantum fluctuations of the impurity due to the coupling
to the Goldstone modes. Since the operators $S_a$ obey the
spin algebra they affect the impurity spin by inducing
transitions between the eigenstates of $S_3$.
The problem can be reduced to a one-dimension (1D) problem using
an s-wave expansion \cite{leggett} with a
%(in analogy with the
%fermionic waves of the Kondo problem). After a simple algebra
%one can show that the problem at hand can be described by the
Hamiltonian, $H=H_0+H_I \delta(x)$, where:
\begin{eqnarray}
H_0 &=& \frac{1}{2} \sum_{a=1,2} \int_0^{\infty} dx \left[\Pi_a^2(x)+ 
c^2 (\partial_x \Phi_a(x))^2\right] \, ,
\nonumber
\\
H_I &=&  \Delta S_3 + \sqrt{\frac{g}{(2 \pi)^d}}
\sum_a \lambda_a \int_0^{\infty} dk
k^{(d-1)/2} \, \, \Phi_a(k) S_a \, ,
\label{hamil}
\end{eqnarray}
where $\Phi_a(x)$ and $\Pi_a(x)$ with $a=1,2$ 
are conjugate real scalar fields on the half line ($\Phi_a(k)$
is the Fourier transform of $\Phi_a(x)$ where $k$ is the
momentum). It is convenient to further decomposed the fields
into right, $\Phi_{a,R}(x)$, and left, $\Phi_{a,L}(x)$, moving components
associated with outgoing and incoming waves out of the impurity, 
respectively.
%($[\Phi_a(x),\Pi_b(y)]=i \delta_{a,b} \delta(x-y)$).
%Furthermore, the fields are subject to the boundary conditions that
%$\Phi_a(x=0)=0$.
Notice that the coupling of the impurity spin to the bosonic reservoir
depends on the dimensionality. This should
be contrasted with the Kondo problem where the coupling to fermions
with a Fermi surface make the problem insensitive to $d$. 
To make contact
with the Kondo problem we return to the path integral language and
trace over the bosonic modes in order to obtain the effective action
${\cal S}_{eff}={\cal S}_{WZ}+{\cal S}_I$ for the impurity alone, where:
\begin{eqnarray}
{\cal S}_I = \int d\tau \Delta S_3(\tau) - \sum_{a=1,2} \gamma_a 
\int d\tau \int d\tau' \frac{S_a(\tau) S_a(\tau')}{|\tau-\tau'|^{\alpha}} \, ,
\label{si}
\end{eqnarray}
where $\gamma_a = (\lambda_a^2 g S_d \Gamma(d-1))/(4 (2 \pi)^d c^d)$
($S_d$ is the area of the hypersphere in $d$ dimensions and $\Gamma(x)$
is the Gamma function) and $\alpha=d-1$ (for a ferromagnet $\alpha=d/2$). 

There is a few well understood 
limits of this problem. Consider first the case where 
$\gamma_1 \neq 0$  but $\gamma_2 = 0$ and $\alpha=2$ ($d=3$). 
In this case the action (\ref{si}) can be mapped 
%into the one obtained by Anderson, Yuval and Hamman 
%\cite{ayh} for
onto the anisotropic Kondo problem \cite{ayh} and is equivalent
to the problem of a classical 1D spin chain with long range interactions 
in a magnetic field. There is a Kosterlizt-Thouless (KT) 
phase transition at $\gamma_1 = 1$: for $\gamma_1<1$, $\Delta$
scales to infinity
indicating that the impurity aligns with the bulk - in the Kondo
language this is equivalent to the formation of the Kondo singlet; 
for $\gamma_1>1$, $\Delta$ is irrelevant under the RG and scales to 
zero - this is the equivalent of the Kondo problem with ferromagnetic 
coupling. The second case is $1<\alpha<2$ and
$\gamma_1=\gamma_2=\gamma$. Notice, on the one hand, 
that in terms of its Fourier transform the second term in (\ref{si}) behaves 
like $|\omega|^{\alpha-1} |S_a(\omega)|^2$ where $\omega$ is the frequency. 
On the other hand, ${\cal S}_{WZ}$ describes the area 
of the unit sphere bounded by the trajectory parameterized by ${\bf
  S}(\tau)$. For a variation of ${\bf S}$ by $\delta {\bf S}$ the
variation in this term is simply $\delta {\cal S}_{WZ} = \delta {\bf S} 
\cdot ({\bf S} \times \partial_{\tau} {\bf S})$ and therefore ${\cal S}_{WZ}$
scales like $\omega$. 
Thus, for $\alpha<2$ ($d<3$) the 
long range interaction is relevant at low energies and one can
disregard ${\cal S}_{WZ}$, that is, the impurity behaves
classically. It may be surprising that as one lower the dimensionality
(decreases $\alpha$) 
the magnetic impurity behaves classically but this results
from the fact that the interactions in imaginary time become longer ranged. 
%Notice that the origin of this interaction has to do with the emission
%of a boson at time $\tau'$ and its absorption at time $\tau$. 
%As one lowers the dimensionality the phase space is reduced and the time
%of return of the bosons is also reduced, increasing the range of interaction. 
By disregarding ${\cal S}_{WZ}$ the problem reduces to a classical XY chain
with long-range interactions \cite{kost} 
in the presence of a field in the Z direction
(proportional to $\Delta$) where $\gamma$ plays the role of the
inverse of the temperature. $\Delta$ is a relevant
perturbation and the spin always orders with the bulk without quantum effects. 
%When $\Delta=0$ this problem has a phase transition
%at finite coupling $\gamma_c = (\alpha-1)/(2 (2-\alpha))$: for
%$\gamma > \gamma_c$ the spin orders in the XY plane while for
%$\gamma < \gamma_c$ $\gamma$ scales to zero under the RG indicating
%that the spin rotates freely. 
For the $d=2$ QHAF this problem has been studied
via spin-wave T-matrix scattering 
\cite{sasha}, field theoretical methods
\cite{sachdev}, and quantum Monte Carlo \cite{sandvik}.
Finally, when $\alpha>2$ ($d>3$) the second
term in (\ref{si}) is irrelevant and the
spin effectively decouples from the fluctuations of the 
environment at low energies, 
that is, the problem can be described in terms of 
a quantum spin in the Weiss molecular field alone. 

It is clear that the case of most interest is when $\alpha=2$ 
and $\gamma_1,\gamma_2 \neq 0$. Notice that for an antiferromagnet
this implies $d=3$ which is also the case of experimental interest.
Returning to (\ref{hamil}), it is convenient to work with a single
field on the entire real line by the unfolding transformation:
$\phi_L(x)=\Phi_{1,L}(x)$, $\phi_R(x)=\Phi_{2,R}(x)$, and
$\phi_L(-x)=\Phi_{1,R}(x)$, $\phi_R(-x) = \Phi_{2,L}(x)$, all
for $x>0$. In this case 
the Hamiltonian can be written as (we set $c=1$):
\begin{eqnarray}
H &=& \int_{-\infty}^{+\infty} dx \sum_{\alpha=R,L}
\left(\partial_x \phi_{\alpha}(x)\right)^2 
+ \delta(x) \left[ \Delta S_3\right. 
\nonumber
\\
&-& \left. \sqrt{8 \pi} \kappa_1 \partial_x \phi_R(x=0) S_1
- \sqrt{8 \pi} \kappa_2 \partial_y \phi_L(x=0) S_2 \right] \, ,
\label{hamil_line}
\end{eqnarray}
where $\phi_R(x)$ and
$\phi_L(x)$ are right and left moving fields, and
$\kappa_{1,2} = g^{1/2} \lambda_{1,2}/(4 \pi^{7/2})$
are the new couplings. Similar problems  
to the one described by (\ref{hamil_line}) have
been studied in the past. In one case a single
heat bath is coupled to different spin components but
in the absence of the Weiss field \cite{sengupta,zimanyi},
in another, the Weiss field is considered only in 
first order \cite{gergely}. In our case it is not only important
to have two independent heat baths but we also consider
the strong renormalization of the Weiss field.

\begin{figure}
\epsfysize5 cm
\hspace{1cm}
\epsfbox{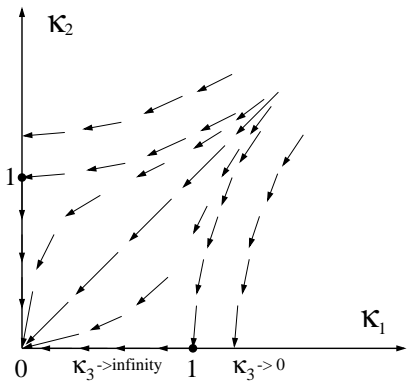}
\caption{RG flow in the $\kappa_1 \times \kappa_2$ plane.}
\label{rgflow}
\end{figure}

We have performed RG calculations
for (\ref{hamil_line}) in the Coulomb gas formulation 
\cite{cg} in two different
limits: ({\it i}) $\kappa_{1,2} \ll 1$; ({\it ii}) 
$\kappa_1 \ll 1$ with $\kappa_2$ arbitrary and vice-versa. 
The RG equations are:
\begin{eqnarray}
\partial_{\ell} \kappa_1 &=& -\kappa_1 \kappa_2^2 - \kappa_1 \kappa_3^2 \, ,
\nonumber
\\
\partial_{\ell} \kappa_2 &=& - \kappa_2 \kappa_1^2 - \kappa_2 \kappa_3^2 \, ,
\nonumber
\\
\partial_{\ell} \kappa_3 &=& (1-\kappa_1^2 -\kappa_2^2) \kappa_3 \, ,
\label{rgeq}
\end{eqnarray}
where $\kappa_3 = \Delta/D_0 \ll 1$ and $\ell = \ln(D/D_0)$.
Notice that, as expected, these equations are symmetric under
the exchange of couplings $\kappa_1$ and $\kappa_2$.
There are 3 fixed points: (1) the trivial one at 
$\kappa_1=\kappa_2=\kappa_3=0$, (2) $\kappa_1=\kappa_3=0$, $\kappa_2=1$,
and (3) $\kappa_2=\kappa_3=0$, $\kappa_1=1$. The two non-trivial
fixed points are the KT transitions of the anisotropic Kondo model.
In the $\kappa_1 \times \kappa_2$ plane ($\kappa_3 \neq 0$) 
the RG flow is shown in Fig.\ref{rgflow}. 

Notice that when the couplings are large 
and anisotropic ($\kappa_1 \gg \kappa_2>1$)
the RG indicates that one of the couplings
flow a fixed value while the others flow to zero ($\kappa_3 \to 0$). 
In terms of (\ref{hamil_line}) this indicates that the impurity spin aligns in
a direction perpendicular to the molecular field which was assumed
to point in the Z-direction in (\ref{action_order}). In the Kondo
language this is the equivalent to the Kondo effect with ferromagnetic
coupling, when the impurity decouples from the environment. 
A possibility that is not considered in this work is associated
with the formation of a spin texture around the impurity spin.
In a classical spin system a spin texture can be formed in the
bulk spins due to the presence of strong and/or anisotropic interactions. The   
spin texture can follow the impurity as it tunnels
invalidating the methods used here (an instanton
calculation is required to take into account the collective nature
of the texture) \cite{next}. Our results are only valid if
no spin texture is formed around the magnetic impurity.

In the isotropic case when $\kappa_1=\kappa_2 =\kappa$ the
RG equations become:
\begin{eqnarray}
\partial_{\ell} \kappa &=& - \kappa^3 - \kappa \kappa_3^2 \, ,
\nonumber
\\
\partial_{\ell} \kappa_3 &=& (1- 2 \kappa^2) \kappa_3 \, .
\label{rg_iso}
\end{eqnarray}
Observe that contrary
to the KT transition $\kappa_3(\ell)$ always scales
towards strong coupling indicating the relevance of the
molecular field (although it decreases initially under the RG
if $\kappa>1/\sqrt{2}$). 
However, the RG breaks down at a scale $\ell^* = \ln(D_0/T_A)$
when $\kappa_3(\ell^*) \approx 1$. $T_A$ is the crossover energy 
scale from weak to strong coupling (the equivalent of the Kondo temperature).
It is easy to see that the value of $T_A$ depends strongly on the bare value
of $\kappa(\ell=0)$.
If $\kappa(0) < \kappa_3(0)$ the
$\kappa^3$ term in (\ref{rg_iso}) does not play a role, the flow is 
essentially the same as the usual KT flow and 
$T_A \approx  D_0 [\kappa_3(0)]^{1/(1-2 \kappa^2(0))} \approx \Delta [1-
2 \kappa^2(0) \ln(D_0/\Delta)]$. 
If, on the other hand, $\kappa(0) > \kappa_3(0)$ then the 
$\kappa^3$ term dominates and $\kappa_3(\ell)$ flows to strong coupling leading to:
$T_A \approx \Delta (1+2 \kappa^2(0) \ln(D_0/\Delta))^{-1}$.
We immediately notice that the $\kappa^3$ term in the
RG destroys the KT transition. Unlike the Kondo problem the system 
retains coherence even at large coupling and is never overdamped.
This is a quantum mechanical effect and
comes from the fact that the spin operators do not commute. While
the $S_1$ operator in (\ref{hamil_line}) wants to orient the impurity
spin in its direction, the same happens for the $S_2$ operator. In a
classical system (large $S$) the spin would orient in a finite angle
in the XY plane. However for a $S=1/2$ impurity this is not possible
and the impurity coupling is effectively {\it quantum frustrated} reducing
the effective coupling to the environment. Another interesting feature of
the RG flow is that for $1>\kappa(0)>1/\sqrt{2}$ the value of
$\kappa^*=\kappa(\ell^*) \approx \ln(D_0/T_A)/2$ is essentially independent of $\kappa(0)$
at energy scale $T_A$. While $T_A$ gives the crossover energy scale
between weak and strong coupling, $\kappa^*$ provides information about the
dissipation rate, $\tau^{-1}$, of the impurity dynamics. Our results indicate that 
for $\kappa(0)$ sufficiently large, $\tau^{-1}$ is
independent of the initial coupling to the bosonic baths.
\begin{figure}
\epsfysize5.5 cm
\hspace{1cm}
\epsfbox{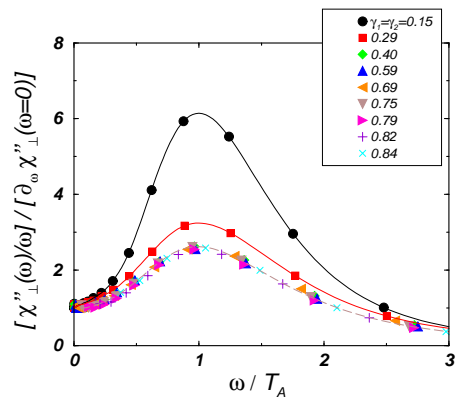}
\caption{$\chi^{''}_{\perp}(\omega,\alpha)/\omega$ as a function of $\omega/T_A$.}
\label{colapse}
\end{figure}

\begin{figure}
\epsfysize5 cm
\hspace{1cm}
\epsfbox{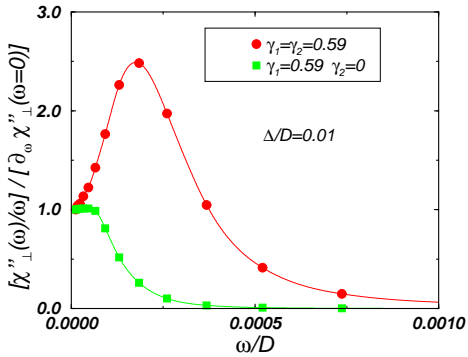}
\caption{$\chi^{''}_{\perp}(\omega)/\omega$ as a function of $\omega/D$.}  
\label{comparison}
\end{figure}

In order to investigate the dynamical correlations 
we study the frequency dependent impurity spin correlation function, 
$\chi^{''}_{\perp}(\omega)=  \Im\left\{\int_0^{\infty} dt e^{i \omega t} 
\langle [S_1(t),S_1(0)]\rangle\right\}$. 
One of the great advantages of writing the problem in the
form of an impurity problem as in (\ref{hamil_line}) is that it 
can be accurately studied by numerical renormalization group 
methods (NRG) \cite{costi}. In order to perform the NRG calculation
we transform (\ref{hamil_line}) into a 1D fermionic problem 
and map the bosonic couplings
onto fermionic coupling by studying the spectrum of both problems \cite{next}. 
In Fig.\ref{colapse} we plot 
$\chi^{''}_{\perp}(\omega)/\omega$ for different values of 
$\gamma_1=\gamma_2=\gamma$ as a function of $\omega/T_A$.
We see that the curves collapse for $\gamma>0.4$ (quasi-scaling) 
while deviations are observed for small enough
$\gamma$. This result agrees with the RG since the width of
$\chi^{''}_{\perp}(\omega)/\omega$ is exactly the dissipation rate, $\tau^{-1}$, which
becomes independent of $\gamma$ (or $\kappa(0)$) for $\gamma > 0.4$. 
In Fig.\ref{comparison} we compare  the behavior of the correlation function
in the isotropic case ($\gamma_1=\gamma_2=0.59$) to the behavior
in the strongly anisotropic case, that is, in the DTLS 
($\gamma_1=0.59$ and $\gamma_2=0$). The differences are striking.
While in the isotropic case the peak in the response at $\omega = T_A$ 
remains, that is the system is underdamped, it has disappeared in the
anisotropic case where the dynamics is overdamped. 

Since the RG indicates that the transverse couplings of the impurity to the
environment always flow to $\kappa \ll 1$ one could use perturbation theory
to calculate $\chi^{''}_{\perp}(\omega)$. However, perturbation theory in
$\kappa$ only generates a Dirac delta peak at $\omega=\Delta$, that is, 
$\tau^{-1}=0$ . 
In order to get a finite $\tau^{-1}$ one needs a non-perturbative
calculation. The RPA with the bare parameters replaced by the renormalized 
ones gives \cite{next}:
\begin{eqnarray}
\frac{\chi^{''}_{\perp}(\omega)}{\omega} = 
\frac{(\pi/2) [\arctan(\Delta \tau)]^{-1} T_A/\tau}{
\left(\omega^2-(T_A)^2-1/\tau^{2}\right)^2+4 
\omega^2/\tau^2} \, ,
\label{rpa}
\end{eqnarray}
where $\tau^{-1} \approx (\kappa^*)^2 T_A$. Notice that (\ref{rpa}) 
reduces to a Dirac delta function at $\omega=\Delta$ as $\kappa(0) \to 0$, 
as expected. We find that this approximation is good for $\omega \ll
T_A$ and also describes well the NRG results for all $\omega <D_0$
when $D_0>T_A \gg D_0 \kappa^*$.
In the zero frequency limit (\ref{rpa}) reduces to 
$\chi_{\perp}(\omega=0) \approx (\kappa^*)^2 \omega/(T_A)^2 + {\cal
  O}((\kappa^*)^4)$ 
and the Kramers-Kronig relation immediately leads to 
$\chi_{\perp}(\omega=0,T=0) = \pi/[8 (1+(\kappa^*)^4)
\arctan(1/(\kappa^*)^2)] 1/T_A \approx 1/(4 T_A) + {\cal O}((\kappa^*)^2)$. 
For $D_0>\omega \gg T_A > D_0 \kappa^* $ (\ref{rpa})
agrees with the NRG results giving 
$\chi^{''}_{\perp}(\omega) \propto 1/\omega^3$. 
In the case where $T_A \ll D_0 \kappa^*$, in the frequency and temperature
range: $T_A \ll \omega,T \ll D_0$, we find:
$\chi^{''}_{\perp}(\omega) \approx \pi/[8 \kappa^2(0) \omega \ln^2(D_0/\omega)]$ 
and $\chi_{\perp}(T) \approx 1/[8 \kappa^2(0) T \ln(D_0/T)]$ \cite{next}.

In summary, we have studied a problem of a spin $1/2$ quantum impurity
coupled to the Goldstone modes of a magnetically ordered system and found
that the problem maps into a generalization of the DTLS that shows
no decoherence even in strong coupling. We have calculated the frequency
and temperature
behavior of the impurity susceptibility that can be measured directly
in a NMR experiment. We assign the destruction of decoherence (and the KT
transition) to a quantum frustration between non-commuting spin operators.
This result may have implications in quantum computation where
decoherence effects are detrimental and the use of quantum frustration
may be explored as a way to avoid decoherence.

We are grateful to P.~Carreta, W.~Buyers, A.~Chernyshev, T.~Costi, B.~Halperin, 
D.~MacLaughlin, A.~Sandvik, S.~Sachdev, M.~Silva Neto, and M.~Vojta, for illuminating conversations.
G.~Z. acknowledges support from grants OTKA F030041 and NSF-OTKA INT-0130446.
I.~A. acknowledges NSF DMR-0203159 for support.


\begin{references}

\bibitem{leggett}A.~J.~Leggett {\it et al}, Rev.~Mod.~Phys. {\bf 59}, 1 (1987).
\bibitem{bethe-ansatz}A.~M.~Tsvelik, and P.~G.~Wiegmann, Adv.~Phys. {\bf 32}, 453 (1983).
\bibitem{ayh}P.~W.~Anderson {\it et al}, Phys.~Rev.~B {\bf 1}, 4464 (1970).
\bibitem{cft}I.~Affleck, Acta Phys.~Polon. B {\bf 26}, 1869 (1995).
\bibitem{saleur}F.~Lesage {\it et al}, Phys.~Rev.~Lett. {\bf 76}, 3388 (1996).
\bibitem{sengupta}J.~L.~Smith, and Q.~Si, Europhys.~Lett. {\bf 45} 228 (1999);
A.~Sengupta, Phys.~Rev.~B {\bf 61}, 4041 (2000).
\bibitem{ferro}We focus on the QHAF but ferromagnetism is analogous.
\bibitem{next}E.~Novais {\it et al.}, unpublished.
\bibitem{buyers}R.~A.~Cowley, and W.~J.~L.~Buyers, Rev.~Mod.~Phys. {\bf 44},
  406 (1972).
\bibitem{sachdev}S.~Sachdev, and M.~Vojta, cond-mat/0303001.
\bibitem{chn}S.~Chakravarty {\it et al}, Phys.~Rev.~B {\bf 39}, 2344 (1989).
\bibitem{kost}J.~M.~Kosterlitz, Phys.~Rev.~Lett. {\bf 37}, 1577 (1976).
\bibitem{sasha}A.~L.~Chernyshev {\it et al}, Phys.~Rev.~B {\bf 65}, 104407 (2002).
\bibitem{sandvik}K.~H.~Hoglund, and A.~W.~Sandvik, cond-mat/0302273.
\bibitem{zimanyi}A.~Zawadowski, and G.~T.~Zim\'anyi, 
Phys.~Rev.~B {\bf 32}, 1373 (1985).
\bibitem{gergely}L.~Zhu, and Q.~Si, Phys.~Rev.~B {\bf 66}, 024426 (2002);
G.~Z\'arand, and E.~Demler, Phys.~Rev.~B {\bf 66}, 024427 (2002).
\bibitem{cg}E.~Novais {\it et al}, Phys.~Rev.~B {\bf 66}, 174409 (2002).
\bibitem{costi}K.~G.~Wilson, Rev.~Mod.~Phys. {\bf 47} 773 (1975); T.~A.~Costi
  in ``Density Matrix Renormalization'', eds. I.~Peschel {\it et  al.} (Springer, 1999).




\end{references}
\end{document}